\documentclass[prb,twocolumn,showpacs,amsmath,amssymb]{revtex4}

\usepackage{graphicx}%Include figure files
\usepackage{dcolumn}%Align table columns on decimal point
\usepackage{bm}% bold math
\usepackage{epsfig}

\begin{document}

\title{Transverse Spin-Orbit Force in the Spin Hall Effect in Ballistic Semiconductor Wires}

\author{Branislav K. Nikoli\' c}
\author{Liviu P. Z\^ arbo}
\author{Sven Welack}
\altaffiliation[Present address: ]{Institut f\" ur Physik, Technische Universit\"at, D-09107 Chemnitz, Germany}
\affiliation{Department of Physics and Astronomy, University
of Delaware, Newark, DE 19716-2570, USA}

\begin{abstract}
We introduce the spin and momentum dependent {\em force operator} which is defined by
the Hamiltonian of a {\em clean} semiconductor quantum wire with homogeneous Rashba spin-orbit 
(SO) coupling attached to two ideal (i.e., free of spin and charge interactions) leads.  Its 
expectation value in the spin-polarized electronic wave packet injected through the leads explains 
why the center of the packet gets deflected in the transverse direction. Moreover, the corresponding {\em spin density} will be dragged along the transverse direction to generate an out-of-plane spin accumulation of opposite signs on  the lateral edges of the wire, as expected in the phenomenology of the spin Hall  effect,
when spin-$\uparrow$ and spin-$\downarrow$ polarized packets (mimicking the injection of
conventional unpolarized charge current) propagate  simultaneously  through the wire.
We also demonstrate that spin coherence of the injected spin-polarized wave packet will  gradually
diminish (thereby diminishing the ``force'') along the SO coupled  wire due to the entanglement of
spin and orbital degrees of freedom  of a single electron, even in the absence of any impurity scattering.
\end{abstract}

\maketitle

The classical Hall effect~\cite{classical_hall} is one of the most widely
known phenomena of condensed matter physics because it represents
manifestation of the fundamental concepts of classical electrodynamics---such
as the Lorentz force---in a complicated solid state environment. A
perpendicular magnetic field ${\bf B}$ exerts the Lorentz force ${\bf
F} = q {\bf v} \times {\bf B}$ on current ${\bf I}$ flowing longitudinally
through metallic or semiconductor wire, thereby separating charges in the 
transverse direction. The charges then accumulate  on the lateral edges of 
the wire to produce a transverse ``Hall voltage'' in the direction  
$q {\bf I} \times {\bf B}$. Thus, Hall-effect measurements reveal the nature of  
the current carriers.

Recent optical detection~\cite{kato,wunderlich} of
the accumulation of spin-$\uparrow$ and spin-$\downarrow$ electrons
on the opposite lateral edges of current carrying semiconductor  wires opens new realm
of the  {\em spin Hall effect}. This phenomenon occurs in the
absence of any external magnetic fields. Instead, it requires
the presence  of SO  couplings, which are tiny
relativistic corrections that can, nevertheless, be much stronger
in semiconductors than in vacuum.~\cite{rashba_review} Besides
deepening our fundamental understanding of the role of SO couplings
in solids,~\cite{rashba_review,spintronics} the spin Hall effect offers 
new opportunities in the design of  all-electrical semiconductor spintronic  
devices that do not require ferromagnetic elements or cumbersome-to-control
external magnetic fields.~\cite{spintronics}

While experimental detection of the strong signatures of the spin
Hall effect brings to an end decades of theoretical speculation
for its existence, it is still unclear what spin-dependent 
forces are responsible for the observed spin separation in different 
semiconductor systems. One potential
mechanism---asymmetric scattering of spin-$\uparrow$ and spin-$\downarrow$ electrons
off impurities with SO interaction---was invoked in the 1970s to predict the emergence
of {\em pure} (i.e., not accompanied by charge transport) spin current,
in the transverse direction to the flow of longitudinal unpolarized
charge current, which would deposit spins of opposite signs on the two
lateral edges of  the sample.~\cite{extrinsic} However, it has been
argued~\cite{bernevig} that in systems with weak SO coupling 
and, therefore, no spin-splitting of the energy bands such spin Hall 
effect of the {\em extrinsic} type (which vanishes  in the absence of 
impurities)  is too small to be  observed in  present experiments~\cite{kato} 
(unless it is enhanced by particular mechanisms involving intrinsic SO coupling 
of the bulk crystal~\cite{engel2005a}).

Much of the recent revival of interest in the spin Hall effect has been ignited by the
predictions~\cite{murakami,sinova} for substantially larger
transverse  pure spin Hall current as a response to the longitudinal
electric field in semiconductors with strong SO coupling which
spin-splits energy bands and induces Berry phase correction
to the group velocity of Bloch wave packets.~\cite{wave_packet}
However,  unusual properties of such {\em intrinsic} spin Hall current
in infinite homogeneous systems, which depends on the  whole Fermi sea 
(i.e., it is determined solely by the equilibrium  Fermi-Dirac distribution function 
and spin-split Bloch band structure) and it is not conserved in the bulk due to 
the presence of SO coupling,~\cite{murakami,sinova} have led to arguments  that its non-zero 
value does not correspond to any real transport of spins~\cite{rashba_eq,zhang}  so that 
no spin accumulation near the boundaries and interfaces could be induced by any intrinsic 
mechanism (i.e., in the absence of impurities~\cite{zhang}).

On the other hand, quantum transport analysis of  spin-charge
spatial propagation through {\em clean} semiconductor wires, which
is formulated in terms of genuine nonequilibrium and Fermi
surface quantities  (i.e., conserved spin
currents~\cite{ring_hall,meso_hall,meso_hall_1} and spin
densities~\cite{accumulation}),  predicts that spin Hall
accumulation~\cite{kato,wunderlich}  of opposite signs on its
lateral  edges will emerge due to strong SO coupling within the
wire region.~\cite{accumulation} Such {\em mesoscopic} spin Hall
effect is determined by the processes on the mesoscale set by the spin 
precession length,~\cite{meso_hall,accumulation}  and depends on the whole 
measuring geometry (i.e.,  boundaries, interfaces,  and  the attached
electrodes) due to the effects of  confinement on the dynamics of transported 
spin in the presence of SO couplings in finite-size  semiconductor structures.~\cite{chao,purity}

Thus, to resolve the discrepancy between different
theoretical answers to such fundamental question as---{\em Are SO interaction  
terms in the effective Hamiltonian of a clean spin-split semiconductor 
wire capable of generating the spin Hall like accumulation on its edges?}---it 
is highly desirable to develop a picture of the transverse motion of spin density 
that would be as transparent  as the familiar picture of the transverse drift of charges 
due to the Lorentz force in the classical Hall effect. Here we offer such a picture by 
analyzing the {\em spin-dependent} ``force'', which can be associated   with any SO coupled quantum Hamiltonian, and its effect on the semiclassical dynamics of spin density of  individual electrons that are
injected as spin-polarized wave packets into the Rashba SO coupled clean semiconductor
quantum wire attached to two ideal (i.e., interaction and disorder free) leads.

The effective mass Hamiltonian of the ballistic Rashba quantum wire is given by
\begin{equation}\label{eq:rashba}
\hat{H} = \frac{\hat{\bf p}^2}{2m^*} + \frac{\alpha}{\hbar}\left(\hat{\bm \sigma}\times\hat{\bf p}\right) \cdot {\bf z} + V_{\rm conf}(y),
\end{equation}
where $\hat{\bf p}$ is the momentum operator in 2D space,
$\hat{\bm \sigma}=(\hat{\sigma}^x,\hat{\sigma}^y,\hat{\sigma}^z)$
is the vector of the Pauli spin operators, and  $V_{\rm conf}(y)$
is  the transverse potential confining electrons to a wire of
finite width. We assume that the wire of dimensions $L_x \times
L_y$ is realized using the two-dimensional electron gas (2DEG),
with ${\bf z}$ being the unit vector orthogonal to its plane.
Within the 2DEG, carriers are subjected to the Rashba SO coupling
of strength $\alpha$, which arises due to the structure inversion
asymmetry~\cite{rashba_review} (of the confining potential and
differing band discontinuities at the heterostructure quantum well
interface~\cite{pfeffer}). 

This Hamiltonian generates a spin-dependent force operator which can be
extracted~\cite{schliemann_force,shen_force} within the Heisenberg
picture~\cite{ballentine} as
\begin{eqnarray} \label{eq:force}
\hat{\bf F}_H & = & m^* \frac{d {\bf r}^2_H}{dt^2} = \frac{m^*}{\hbar^2} [\hat{H},[\hat{\bf r}_H,\hat{H}]]  \\
& = & \frac{2 \alpha^2 m^* }{\hbar^3}  (\hat{\bf p}_H \times {\bf
z}) \otimes \hat{\sigma}^z_H  - \frac{d V_{\rm conf}(\hat{y}_H)}{d
\hat{y}_H} {\bf y} \nonumber.
\end{eqnarray}
Here the Heisenberg picture operators carry the time dependence of quantum evolution, i.e., $\hat{\bf p}_H (t) = e^{i\hat{H}t/\hbar} \hat{\bf p} e^{-i\hat{H}t/\hbar}$, $\hat{\sigma}^z_H(t) = e^{i\hat{H}t/\hbar} \hat{\sigma}^z e^{-i\hat{H}t/\hbar}$, and $\hat{y}_H (t) = e^{i\hat{H}t/\hbar} \hat{y} e^{-i\hat{H}t/\hbar}$, where $\hat{\sigma}^z$, $\hat{\bf p}$, and $\hat{y}$ are in the Schr\" odinger picture and, therefore, time-independent.

Since the force operator~\cite{shen_force} depends on spin through
$\hat{\sigma}^z_H$, which is a genuine (internal) quantum degree of freedom,~\cite{ballentine}
it  does not have any classical analog. Its physical meaning (i.e., measurable
predictions) is contained in the quantum-mechanical expectation values, such as  
$\langle \hat{F}_y \rangle (t) = \langle \Psi(t=0) | \hat{F}_H^y (t) | \Psi(t=0) \rangle$  
obtained by acting with the transverse component $\hat{F}_H^y$ of the vector of the force 
operator  $(\hat{F}_H^x,\hat{F}_H^y)$ on the quantum state  $|\Psi(t=0) \rangle$ of an electron. 
While such ``force'' can always be associated with a given quantum Hamiltonian, its usefulness in 
understanding the evolution of quantum systems is limited---the local nature of
the force equation cannot be reconciled with inherent non-locality
of quantum mechanics. For example, if the force ``pushes'' the volume of
a wave function locally, one has to find a new {\em global} wave
function in accord with the boundary conditions at {\rm infinity}
(the same problem remains well-hidden in the Heisenberg picture where time
dependence is carried by the operators while wave functions are
time-independent). Nevertheless, analyzing the dynamics of spin and
probability densities in terms of the action of local forces can be
insightful for particles described by wave packets (whose probability
distribution is small compared to the typical length scale over which the
force varies).~\cite{wave_packet,ballentine}

\begin{figure}
\includegraphics[scale=0.6]{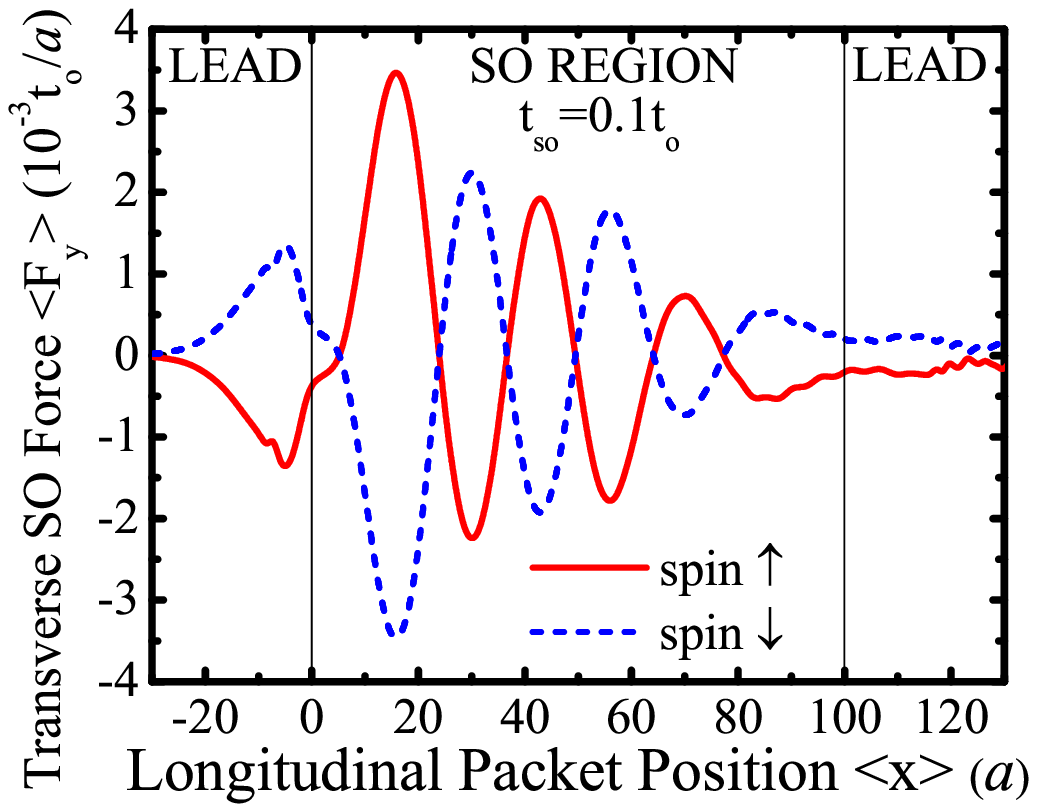}
\includegraphics[scale=0.6]{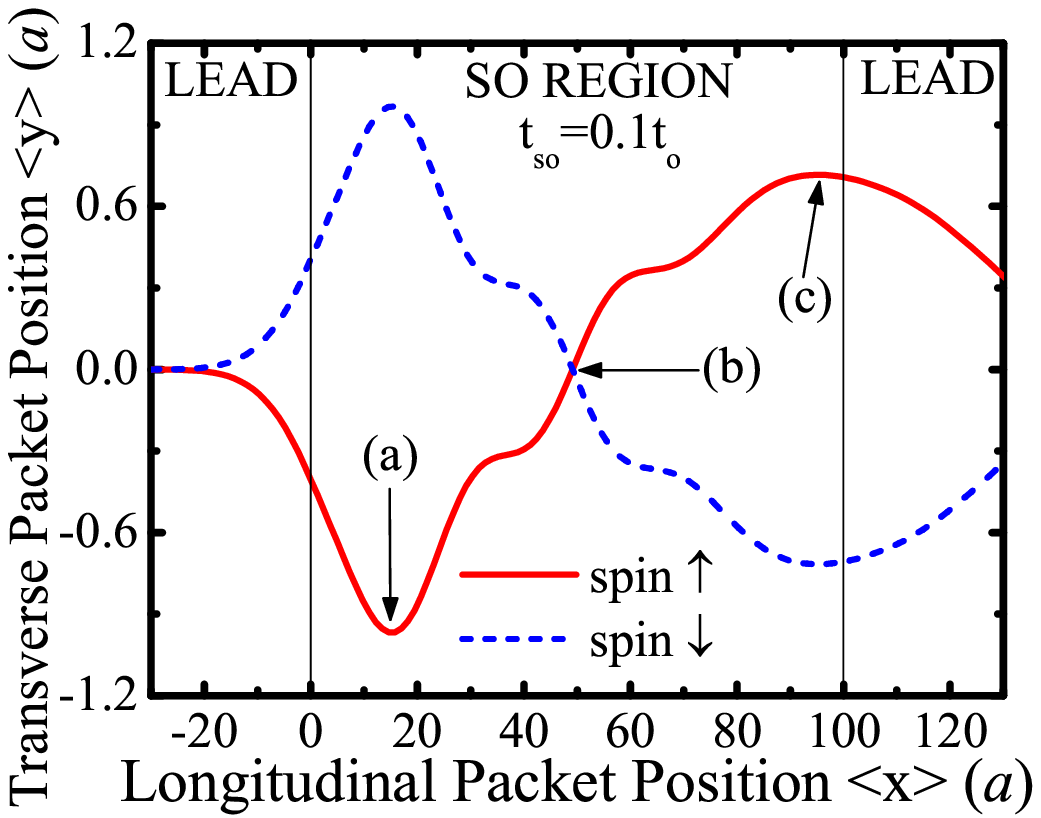}
\includegraphics[scale=0.6]{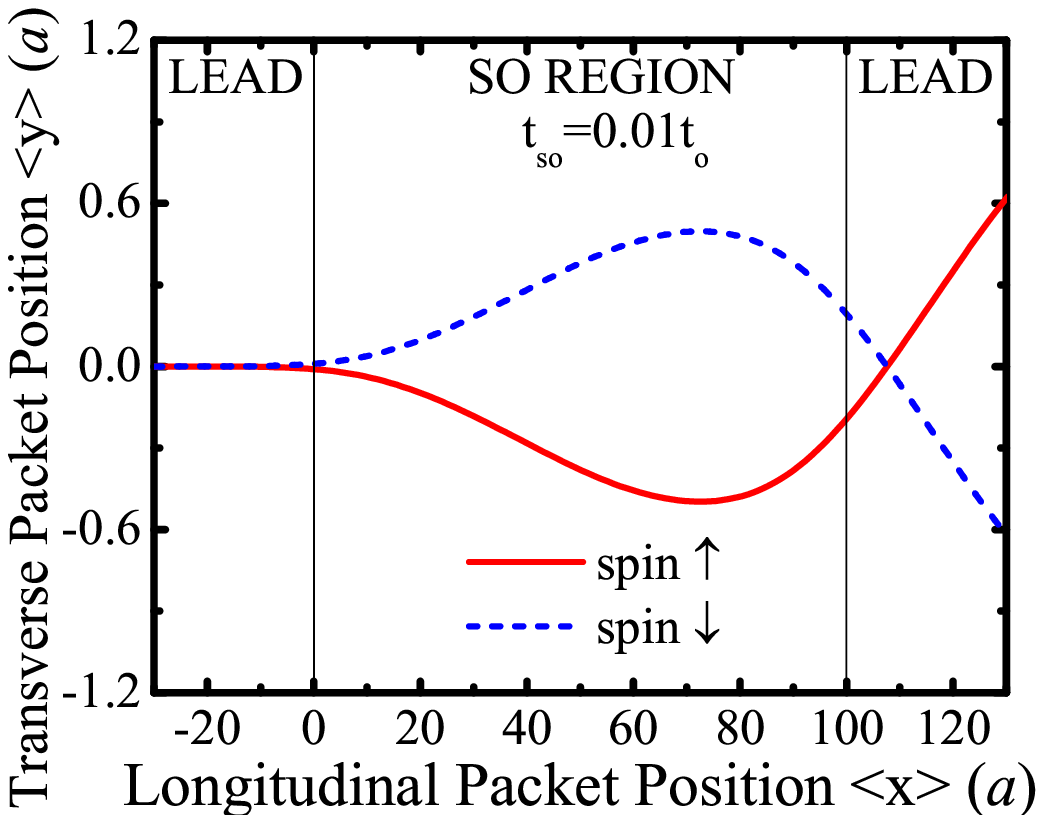}
\caption{(Color online) The expectation value of the transverse component of the SO
force operator (upper panel) in the quantum state of propagating spin wave packet
along the two-probe nanowire. The middle panel shows the corresponding
transverse position of the center of the wave packet  as a function of its
longitudinal  coordinate. The initial state in the  left lead is fully spin-polarized
wave packet Eq.~(\ref{eq:packet}), which is injected into the SO region of the
size $L_x \times L_y \equiv 100a \times 31a$ ($a \simeq 3$ nm) with strong Rashba  
coupling $t_{\rm SO}=\alpha/2a=0.1 t_{\rm o}$ and the corresponding spin precession length 
$L_{\rm SO} = \pi t_{\rm o} a/2t_{\rm SO} \approx 15.7a<L_x$ (middle panel) or weak SO 
coupling $t_{\rm SO} = 0.01 t_{\rm o}$ and $L_{\rm SO}  \approx 157a>L_x$ (lower panel).}
\label{fig:force}
\end{figure}

Therefore,  we examine in Fig.~\ref{fig:force} the transverse SO  
``force'' $\langle \hat{F}_y \rangle$  in the spin wave packet
state, which at $t=0$ resides in the left lead as fully spin-polarized
(along the $z$-axis) and spatially localized wave function~\cite{ohe,serra}
\begin{equation} \label{eq:packet}
\Psi(t=0)  =   C \sin{\left( \frac{\pi y} {(L_y+1)a} \right)} e^{i
k_x x- \delta k_x^2 x^2/4} \otimes \chi_\sigma.
\end{equation}
This is a pure and separable quantum  state $|\Psi(t=0) \rangle = |\Phi\rangle \otimes |\sigma \rangle$  
in the tensor product of the orbital and spin Hilbert spaces ${\mathcal H}_o \otimes {\mathcal H}_s$. 
The orbital factor state $\langle x,y |\Phi \rangle$ consists of the lowest subband of the  hard wall 
transverse confining potential and a Gaussian wave packet along the $x$-axis whose parameters are chosen 
to be $k_xa=0.44$ and $\delta k_xa=0.1$ ($C$ is the normalization constant determined from $\langle \Phi | \Phi \rangle=1$).  The spin factor state is an eigenstate of $\hat{\sigma}^z$, i.e., $\chi_\uparrow  = \left(\begin{array}{c} 1  \\ 0 \end{array}
\right)$ or $\chi_\downarrow  =  \left(
\begin{array}{c} 0  \\ 1 \end{array} \right)$.

Unlike the case~\cite{schliemann_force} of an infinite 2DEG, the exact solutions of the Heisenberg
equation of motion for $\hat{\sigma}^z_H(t)$, $\hat{y}_H(t)$ and $\hat{\bf p}_H(t)$
entering in Eq.~(\ref{eq:force}) are not available for quantum wires of
finite width. Thus, we compute the expectation value $\langle \Psi(t) |
\hat{F}_y | \Psi(t) \rangle$ in the Schr\" odinger picture by
applying the evolution operators $e^{-i\hat{H}t/\hbar}$ present in
Eq.~(\ref{eq:force}) on the wave functions $|\Psi(t) \rangle =
\sum_{n} e^{-iE_n t/\hbar} |E_n \rangle \langle E_n| \Psi(t=0)
\rangle$. To obtain the exact eigenstates~\cite{serra,governale,usaj} $|E_n
\rangle$ and eigenvalues $E_n$, we employ the discretized version
of the Hamiltonian Eq.~(\ref{eq:rashba}). That is, we represent
the Hamiltonian of the Rashba spin-split quantum wire in the basis of
states $|{\bf m} \rangle \otimes |\sigma \rangle$,  where $|{\bf
m} \rangle$ are $s$-orbitals $\langle {\bf r}|{\bf m}\rangle =
\psi({\bf r}-{\bf m})$  located at sites ${\bf m}=(m_x,m_y)$ of
the $L_x \times L_y$ lattice with the lattice spacing $a$
(typically~\cite{purity} $a \simeq 3$ nm). This representation
extracts the two energy scales from the Rashba Hamiltonian
Eq.~(\ref{eq:rashba}): $t_{\rm o}=\hbar^2/(2m^*a^2)$
characterizing hopping between the nearest-neighbor sites without
spin-flip; and $t_{\rm SO}=\alpha/ 2 a$ for the same hopping
process when it involves spin flip.~\cite{accumulation,purity} The
wave vector of the Gaussian packet $k_xa=0.44$ is
chosen~\cite{accumulation} to correspond to the Fermi energy $E_F
= -3.8 t_{\rm o}$ close to the bottom of the band where
tight-binding dispersion relation reduces to the parabolic one of
the Hamiltonian Eq.~(\ref{eq:rashba}). In this representation one
can directly compute the commutators in the definition of the
force operator Eq.~(\ref{eq:force}), thereby bypassing
subtleties which arise when evaluating the transverse component of
the force operator $-dV_{\rm conf}(\hat{y}_H){\bf y}/d\hat{y}_H$
stemming from the hard wall boundary conditions.~\cite{rokhsar}

\begin{figure}
\centerline{\psfig{file=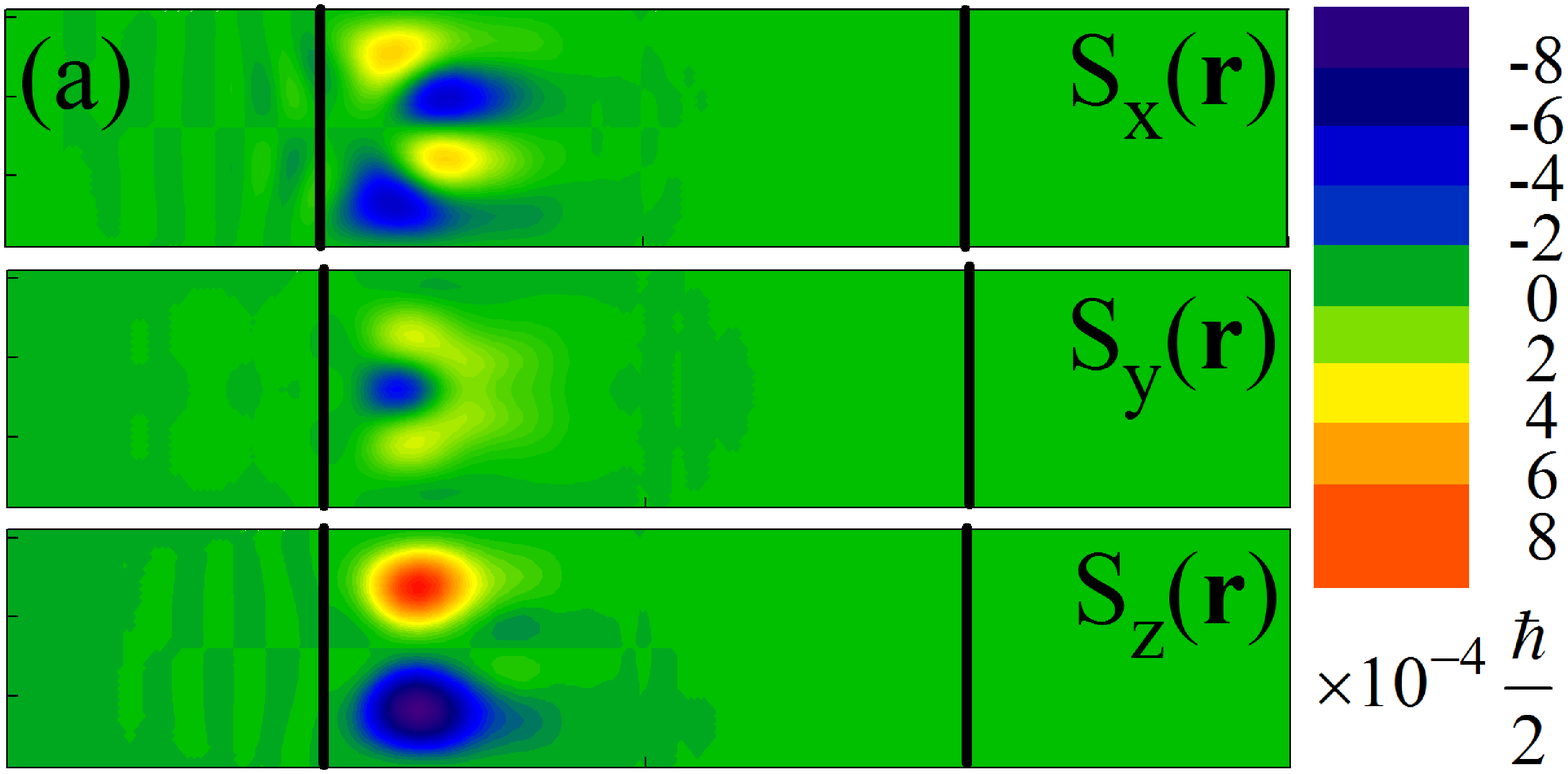,scale=0.25,angle=0}}
\vspace{0.1in}
\centerline{\psfig{file=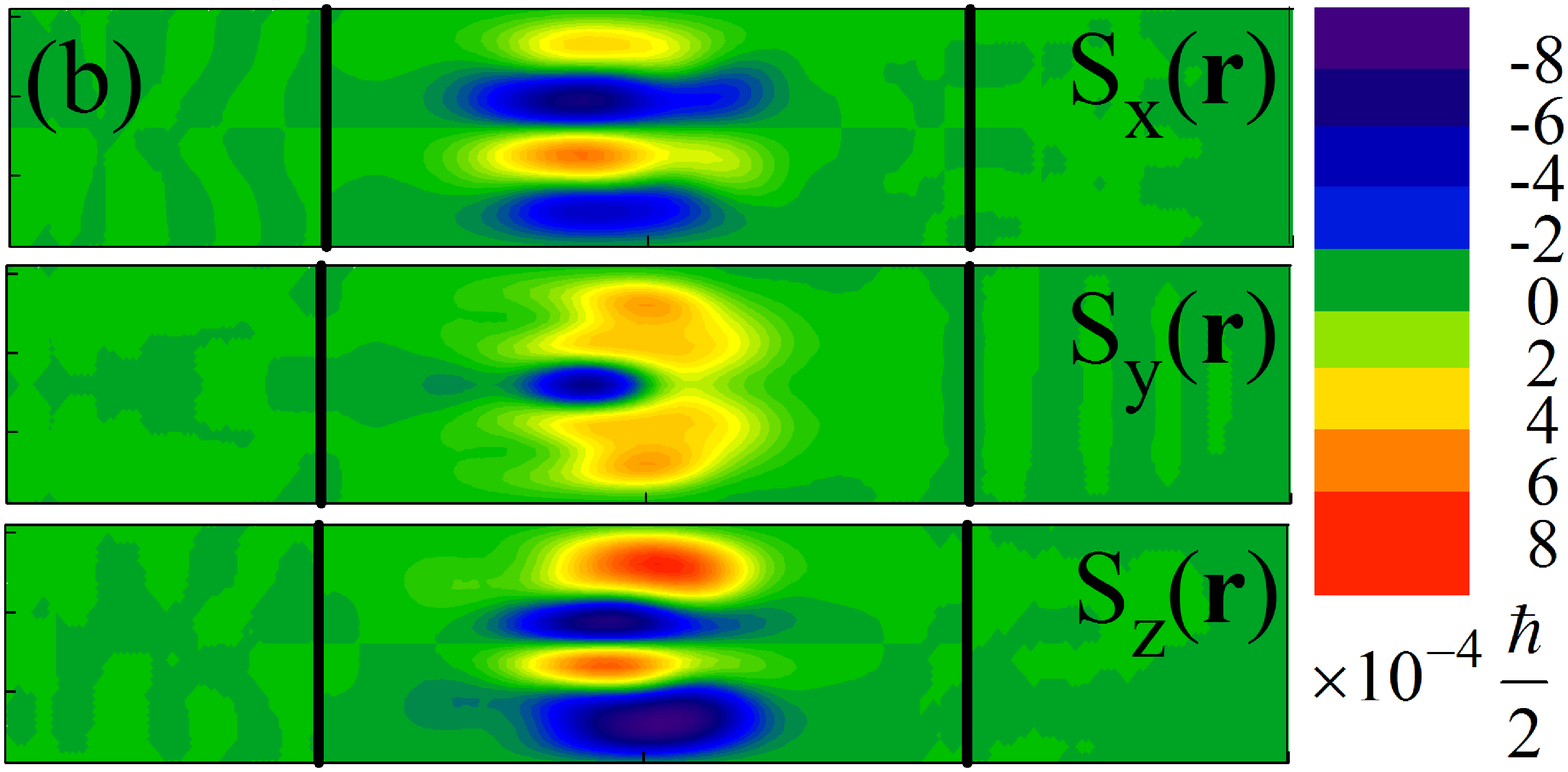,scale=0.25,angle=0}}
\vspace{0.1in}
\centerline{\psfig{file=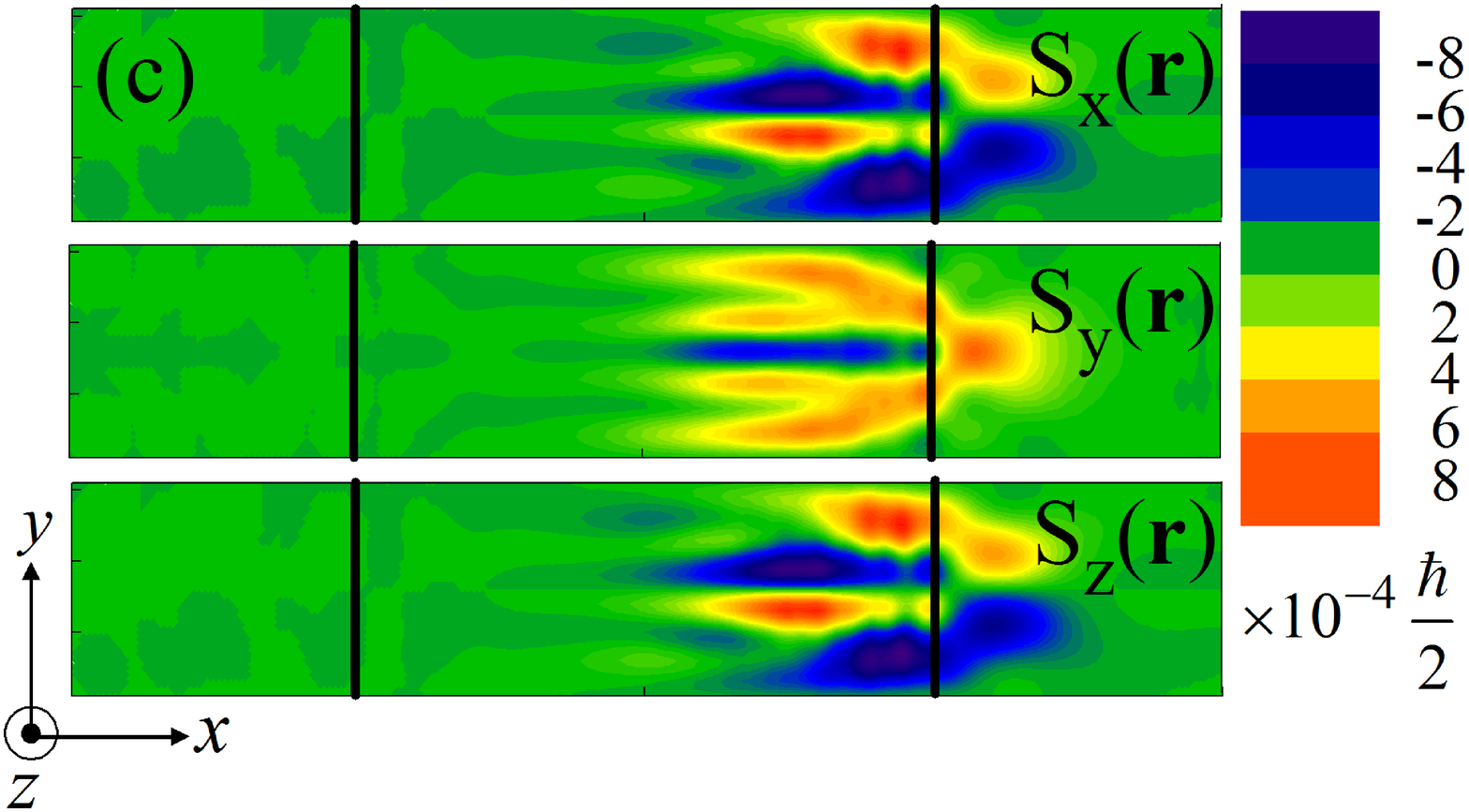,scale=0.25,angle=0}}
\caption{(Color online) The dynamics of spin density ${\bf S}({\bf r}) \equiv
[S_x({\bf r})$, $S_y({\bf r})$, $S_z({\bf r})]$ induced by {\em
simultaneous}  propagation of {\em two} electrons through quantum
wire $100a \times 31a$ with the Rashba SO coupling $t_{\rm
so}=0.1t_{\rm o}$. Both electrons are injected  at $t=0$ from the
left lead, one as spin-$\uparrow$ and the other one as
spin-$\downarrow$ polarized (along the $z$-axis) wave packet
Eq.~(\ref{eq:packet}). The different snapshots of the sum of their
spin densities are taken at the points (a), (b), (c) where the 
transverse SO ``force'' and the $y$-coordinate of  the center of 
these wave packets  have  values shown in the middle panel of
Fig.~\ref{fig:force}.}\label{fig:spin_density}
\end{figure}

Figure~\ref{fig:force} shows that as soon as the front of the
spin-polarized wave packet enters the strongly SO coupled region, its center
$\langle \hat{y} \rangle (t) =  \langle \Psi(t) | \hat{y} |
\Psi(t) \rangle$ will be deflected along the $y$-axis in the same
direction as is the direction of the transverse SO ``force''. However, due to
its inertia the packet does not follow fast oscillations of the SO
``force'' occurring on the scale of the spin precession
length~\cite{accumulation,purity} $L_{\rm SO} = \pi t_{\rm o}a/2t_{\rm SO}$ on which spin
precesses by an angle $\pi$ (note that the spin splitting generates a finite difference of 
the  Fermi momenta, which is the same for all subbands of the quantum wire in the case of 
parabolic  energy-momentum dispersion, so that $L_{\rm SO}$ is equal for all channels~\cite{governale}). 
In contrast  to an infinite 2DEG of the intrinsic spin Hall
effect,~\cite{sinova,rashba_eq,zhang} in quantum wires electron motion
is confined in the transverse direction and the effective momentum-dependent
Rashba magnetic field ${\bf B}_R({\bf k})$ is, therefore, nearly
parallel to this direction.~\cite{governale,purity}  Thus, the change of the direction of the
transverse SO ``force'' is due to the fact that the $z$-axis polarized spin will start
precessing within the SO region since it is not an eigenstate of the Zeeman term
$\hat{\bm \sigma} \cdot {\bf B}_R({\bf k})$ [i.e., of the Rashba term in Eq.~(\ref{eq:rashba})].

The transverse SO ``force'' and the motion of the center of the wave packets
in Fig.~\ref{fig:force} suggests that when {\em two}  electrons with
opposite spin-polarizations are injected {\em simultaneously} into the SO coupled quantum wire
with perfectly homogeneous~\cite{ohe} Rashba coupling, the initially
unpolarized mixed spin state will evolve during propagation through the wire
to develop a non-zero spin density at its lateral edges. This intuitive picture is confirmed
by plotting in Fig.~\ref{fig:spin_density} the spin density,
\begin{eqnarray}\label{eq:spin_density}
{\bf S}_{\bf m}(t) & = & \frac{\hbar}{2} \langle \Psi(t) | \hat{\bm \sigma} \otimes |{\bf m} \rangle \langle {\bf m}| \Psi(t) \rangle \nonumber \\
 & = & \frac{\hbar}{2} \sum_{\sigma,\sigma^\prime} c_{{\bf m},\sigma^\prime}^*(t) c_{{\bf m},\sigma}(t) \langle \sigma^\prime | \hat{\bm \sigma}| \sigma \rangle,
\end{eqnarray}
corresponding to the coherent evolution of two spin wave packets,
$|\Psi (t=0) \rangle = |\Phi \rangle \otimes |\!\! \uparrow \rangle$ and $|\Psi (t=0) \rangle = |\Phi \rangle \otimes |\!\! \downarrow \rangle$, across the wire.

\begin{figure}
\centerline{\psfig{file=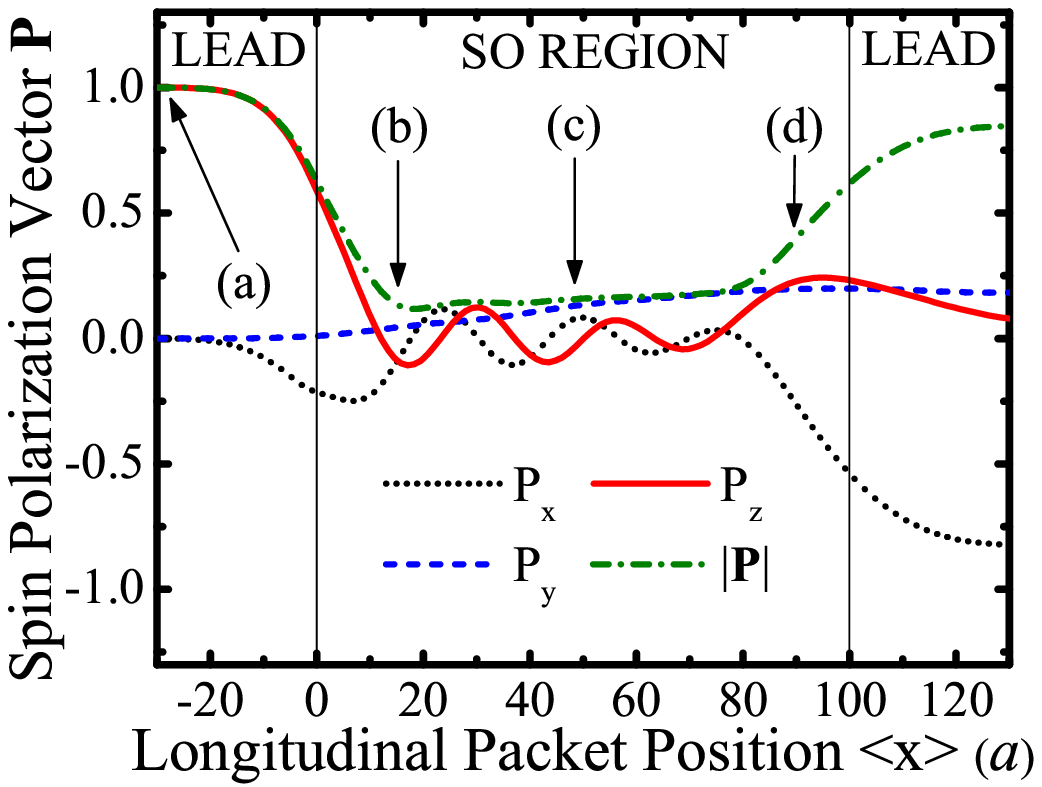,scale=0.65,angle=0}}
\vspace{0.1in}
\centerline{\psfig{file=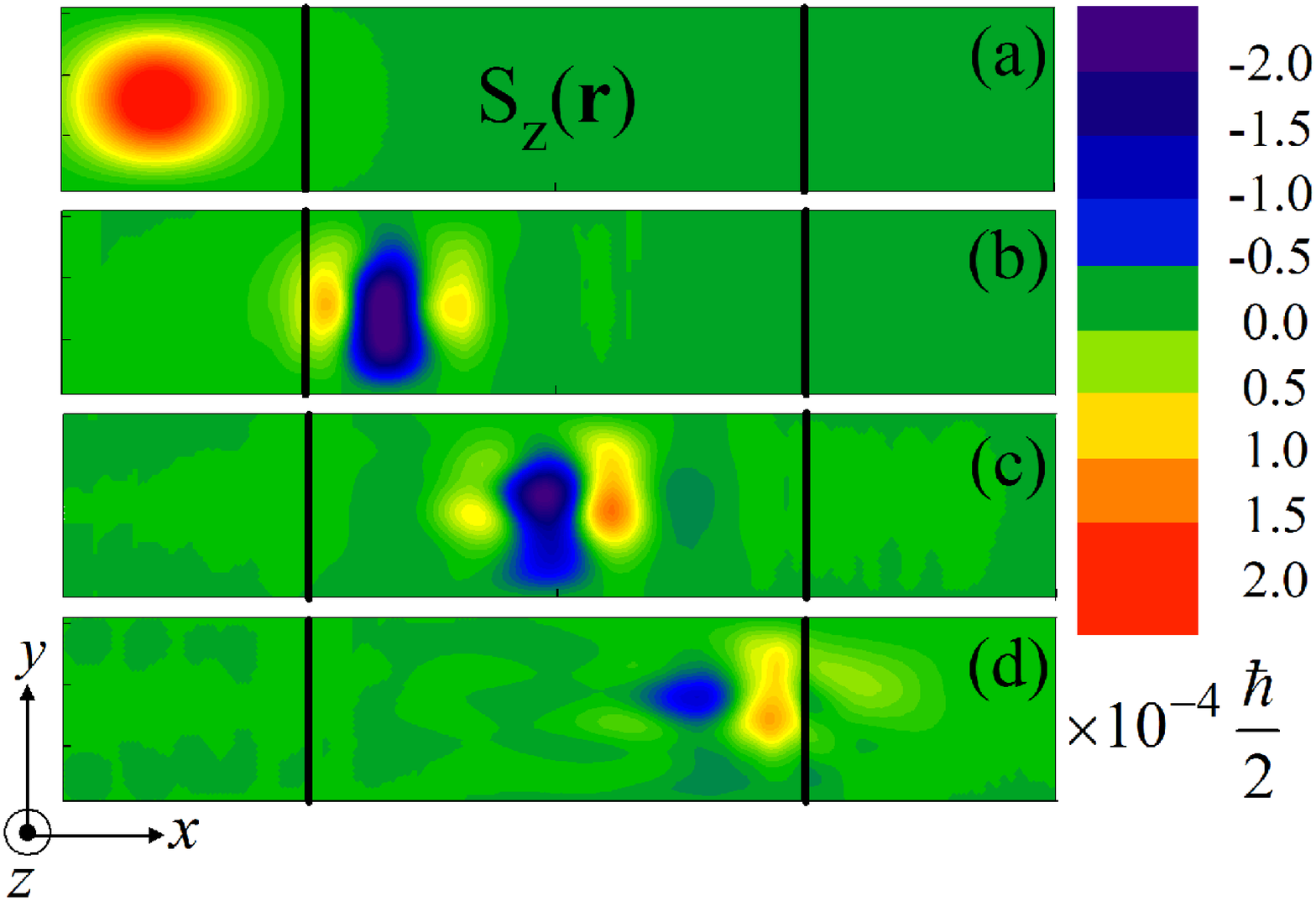,scale=0.23,angle=0}}
\caption{(Color online) Spin precession, as signified by oscillations of the spin
polarization vector $(P_x,P_y,P_z)$, and spin decoherence (as
measured by decrease of the {\em purity} $|{\bf P}|$ below one) of the
spin state of a single electron propagating along the Rashba
quantum wire $100a \times 31a$ with the SO coupling strength
$t_{\rm SO}=0.1t_{\rm o}$ ($L_{\rm SO} \approx 15.7a$). The
electron is injected from the left lead as a spin-$\uparrow$
polarized wave packet, whose spin subsystem is therefore fully
coherent  $|{\bf P}| = 1$ at $t=0$. The bottom panel shows the
$z$-component of the spin density $S_z({\bf r})$ at
different values of $\frac{\hbar}{2} P_z = \int d{\bf r} \, S_z
({\bf r})$ (selected in the upper panel) along the wire.}\label{fig:decoherence}
\end{figure}

The mechanism underlying the decay of the transverse SO ``force'' intensity
is explained in Fig.~\ref{fig:decoherence}, where we demonstrate that
(initially coherent) spin precession is also accompanied by
{\em spin decoherence}.~\cite{purity,galindo}  These  two processes are encoded in the
rotation  of the spin polarization vector ${\bf P}$ and the reduction of its
magnitude  ($|{\bf P}| = 1$ for fully  coherent pure states $\hat{\rho}_s^2 = \hat{\rho}_s$), respectively.
The spin  polarization vector is extracted from the density matrix
$\hat{\rho}_s = (1+{\bf P} \cdot \hat{\bm \sigma})/2$ of the spin subsystem.~\cite{ballentine} The spin density matrix $\hat{\rho}_s$ is obtained as the exact reduced density matrix at each instant of
time by tracing the pure state density matrix $\hat{\rho}(t) = |\Psi (t) \rangle \langle \Psi(t)|$
over the orbital degrees of freedom,
\begin{eqnarray} \label{eq:rho}
\hat{\rho}_s (t) & = & {\rm Tr}_o |\Psi (t) \rangle \langle
\Psi(t)| = \sum_{\bf m} \langle {\bf m} |\Psi (t) \rangle \langle
\Psi(t)|{\bf m}\rangle \nonumber \\ & = & \sum_{\bf m,\sigma,\sigma^\prime}
c_{{\bf m},\sigma}(t) |\sigma \rangle \langle \sigma^\prime|c_{{\bf m},\sigma^\prime}^*(t).
\end{eqnarray}
The dynamics of the spin polarization vector and the spin density shown
in  Fig.~\ref{fig:decoherence} are in one-to-one correspondence
\begin{equation}
\frac{\hbar}{2}{\bf P}(t) = 
      \frac{\hbar}{2} {\rm Tr}_s \,\left[ \hat{\rho}_s(t)  \hat{\bm{\sigma}} \right] = \sum_{\bf m} {\bf S}_{\bf m}(t).
\end{equation}
The incoming quantum state from the left lead in Fig.~\ref{fig:decoherence}
is separable $|\Psi (t=0) \rangle = \sum_{{\bf m},\sigma} c_{{\bf m},\sigma}(t=0)
|{\bf m} \rangle \otimes |\sigma \rangle = |\Phi \rangle \otimes
|\!\! \uparrow \rangle$, and therefore fully spin coherent $|{\bf P}|=1$.
However, in the course of propagation through SO coupled quantum wires it
will coherently evolve into a  {\em non-separable}~\cite{ballentine}
state where spin and orbital  subsystems of the same electron appear to be
{\em entangled}.~\cite{purity,peres} Note that Fig.~\ref{fig:decoherence} also 
shows that at the instant when the center of the wave packet enters the wire 
region, its quantum  state is already highly entangled as quantified by the 
non-zero von Neumann entropy (associated with the reduced density matrix of 
either the spin $\hat{\rho}_s$ or the orbital subsystem $\hat{\rho}_o$)
\begin{eqnarray}\label{eq:entropy}
S(\hat{\rho}_s) = S(\hat{\rho}_o) & = & -\frac{1+|{\bf P}|}{2} \log_2 \left( \frac{ 1+|{\bf P}| }{2} \right) \nonumber \\
&& - \frac{1-|{\bf P}|}{2} \log_2 \left( \frac{1-|{\bf P}|}{2} \right),
\end{eqnarray}
which is a unique measure~\cite{galindo} of the degree of entanglement for  
pure bipartite states (such as  the full state $|\Psi (t) \rangle$ which 
remains pure due to the absence of inelastic processes along the quantum wire).

While this loss of spin coherence (or polarization) is analogous to the well-known DP spin relaxation in
diffusive SO coupled systems,~\cite{spintronics,dp} here the decay of the spin polarization
vector takes place without any scattering  off impurities (or averaging over 
an ensemble of electrons propagating through ballistic SO coupled quantum dot 
structures~\cite{chao}). Instead, it arises due to wave packet spreading (cf. 
lower panel of Fig.~\ref{fig:decoherence}), as well as  due to the presence of 
interfaces~\cite{purity}  (the wave packet is partially reflected at the lead/SO-region 
interface for strong  Rashba coupling) and boundaries~\cite{purity,chao} of the confined 
structure. Thus,  the decoherence mechanism revealed by Fig.~\ref{fig:decoherence} is 
also highly relevant for the  interpretation of experiments on the transport of spin coherence 
in high-mobility semiconductor~\cite{awschalom} and molecular spintronic devices.~\cite{cnt}

The interplay of the oscillating and decaying
(induced by spin precession and spin decoherence, respectively)
transverse SO ``force''  and wave packet inertia leads to spin-$\uparrow$
electron exiting the wire with its center deflected  toward the left
lateral edge and the spin-$\downarrow$ density appearing on the right 
edge~\cite{accumulation} for strong SO coupling $t_{\rm SO}=0.1t_{\rm o}$ in Figs.~\ref{fig:force} and ~\ref{fig:spin_density}.
This picture is only apparently counterintuitive to the na\"ive conclusion
drawn  from  the  form of the  force operator itself Eq.~(\ref{eq:force}), which 
would  suggest that  spin-$\uparrow$ electron is always deflected to the right while moving 
along the Rashba SO  region. While such situation appears in wires shorter 
than  $L_{\rm SO}$ (as shown in the lower panel of Fig.~\ref{fig:force}), in general, 
one  has to take into account the ratio $L_x/L_{\rm SO}$, as well as the strength of the 
SO force $\propto \alpha^2$, to decipher the sign of the spin accumulation on the lateral edges and 
the sign of the corresponding spin currents that will  be pushed into the transverse leads attached 
at those edges.~\cite{meso_hall}

\begin{figure}
\centerline{\psfig{file=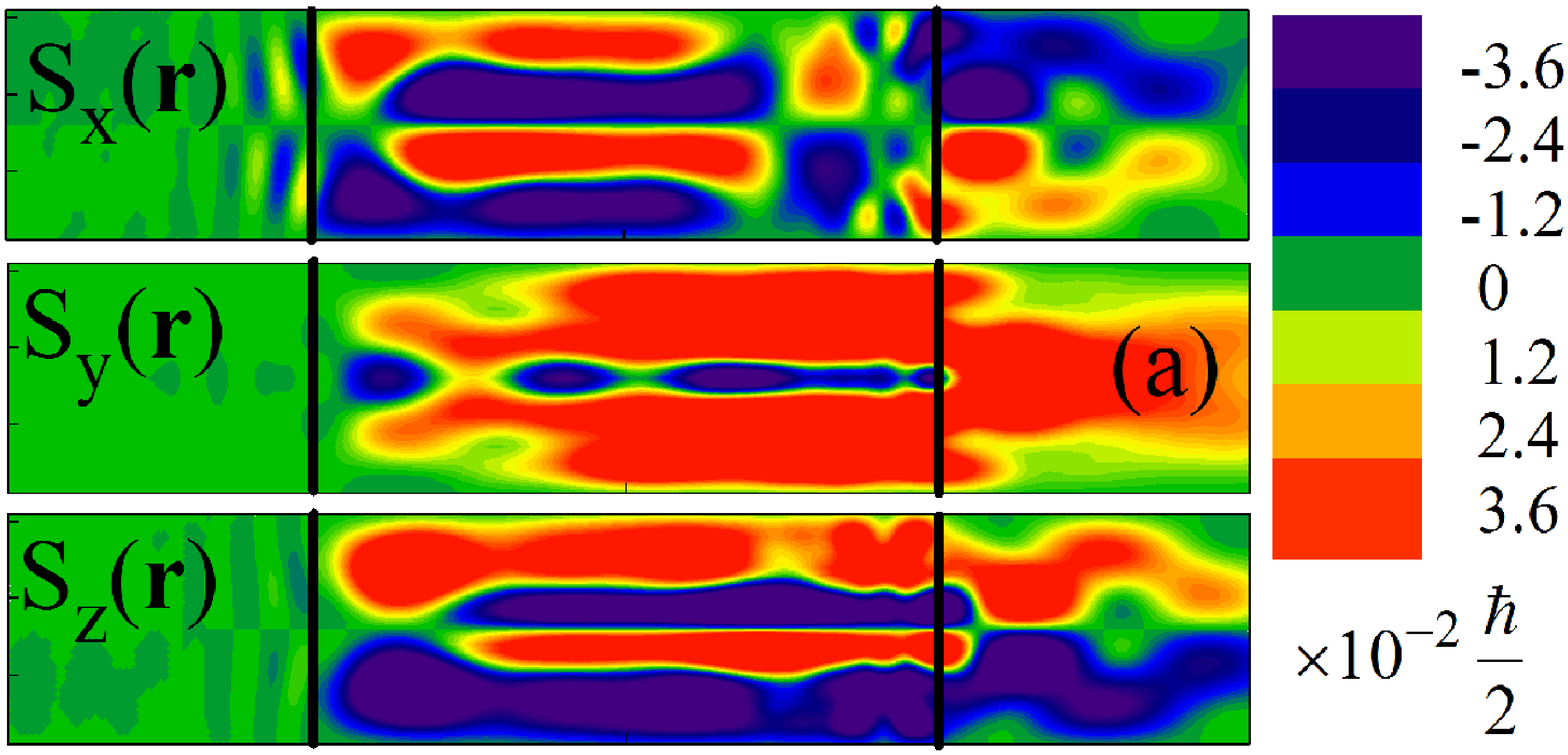,scale=0.25,angle=0}}
\vspace{0.1in}
\centerline{\psfig{file=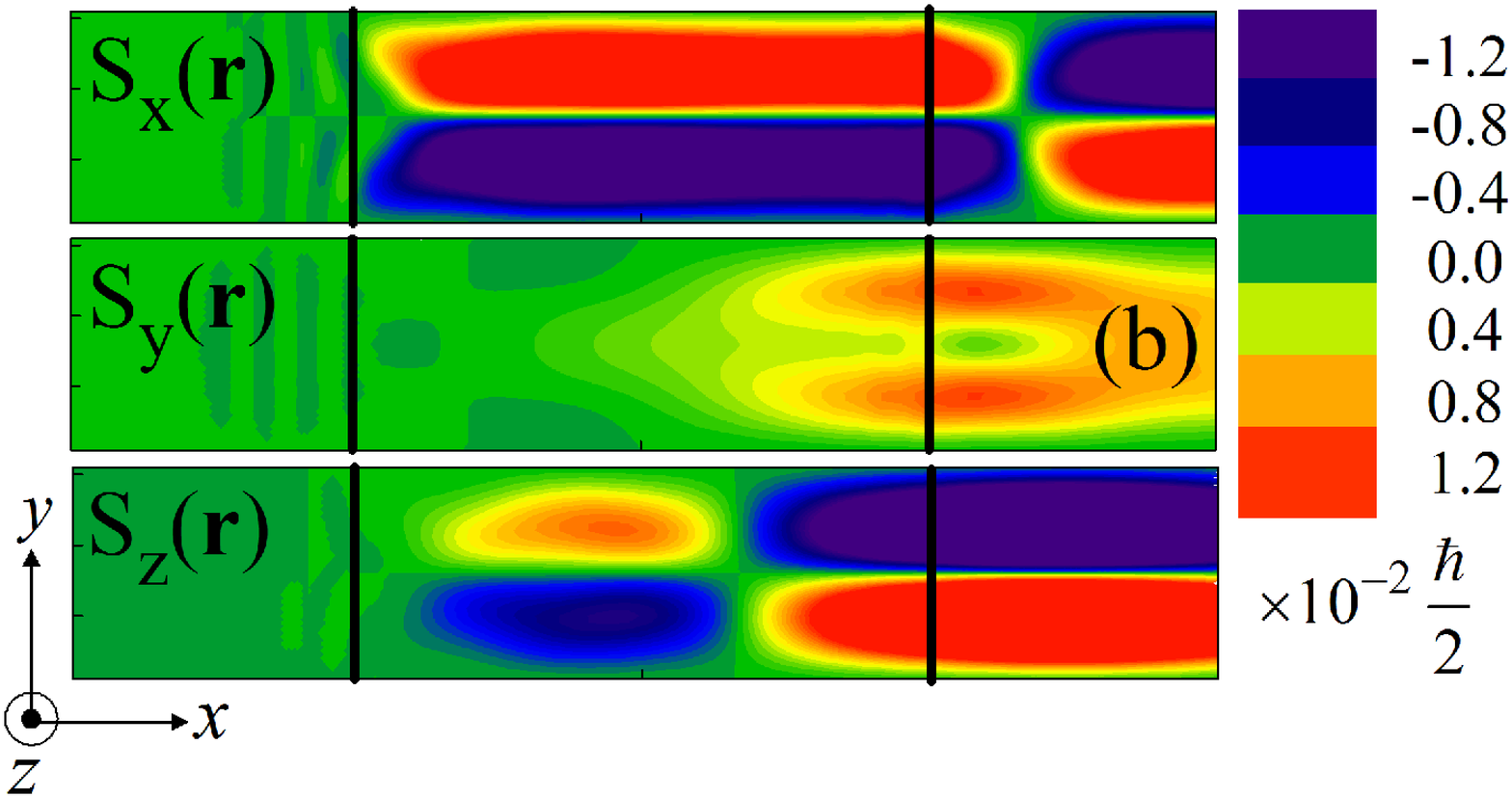,scale=0.23,angle=0}}
\caption{(Color online) The spin accumulation $(S_x({\bf r})$, $S_y({\bf r})$, $S_z({\bf r}))$
induced by the ballistic flow of unpolarized charge current, simulated by
injecting one after  another 600  pairs of spin-$\uparrow$ and
spin-$\downarrow$  polarized (along the $z$-axis)  wave packets from the left
lead, through quantum wire  $100a \times 31a$ with the Rashba SO
couplings: (a) $t_{\rm SO}=0.1t_{\rm o}$ ($L_{\rm SO} \approx 15.7a$) and (b) $t_{\rm SO}=0.01t_{\rm o}$
($L_{\rm SO} \approx 157a$).}\label{fig:she_accumulation}
\end{figure}

When we inject pairs of
spin-$\uparrow$ and spin-$\downarrow$ polarized wave packets one
after another,  thereby simulating the flow of unpolarized ballistic
current through the lead--wire--lead structure (where electron does not
feel any electric field within the clean quantum wire region),~\cite{accumulation}
we find in Fig.~\ref{fig:she_accumulation} that the deflection of the spin densities of individual electrons in
the transverse direction will generate non-zero spin accumulation
components   $S_z({\bf r})$ and $S_x({\bf r})$ of the opposite sign
on the lateral edges of the wire. While recent experiments find $S_z({\bf r})$
with such properties to be the strong signature of the spin Hall
effect,~\cite{kato,wunderlich} here we confirm the conjecture of Ref.~\onlinecite{accumulation}
that $S_x({\bf r})$ can also emerge as a distinctive feature of the mesoscopic spin Hall effect in 
confined Rashba spin-split structures---it arises due to the precession (Fig.~\ref{fig:decoherence}) of transversally deflected  spins. Note that $S_x({\bf r}) \neq 0$ accumulations cannot be  explained by arguments based on the texturelike structure~\cite{governale}  of the spin density of the eigenstates in infinite Rashba quantum wires where~\cite{governale,usaj}  $S_x({\bf r}) \equiv 0$.

In conclusion, the spin-dependent force operator, defined by the
SO coupling terms of the Hamiltonian of a ballistic spin-split
semiconductor quantum wire, will act on the injected spin-polarized
wave packets to deflect spin-$\uparrow$ and spin-$\downarrow$
electrons in the opposite transverse directions. This effect,
combined with precession and decoherence of the deflected spin,
will lead to non-zero $z$- and $x$-components of the spin density
with opposite signs on  the lateral edges of the wire, which represents 
an example of the spin Hall effect phenomenology~\cite{extrinsic,accumulation} that 
has been observed in recent experiments.~\cite{kato,wunderlich} The intuitively appealing
picture of the transverse SO quantum-mechanical force operator (as a counterpart 
of the  classical Lorentz force), which depends on spin through $\hat{\sigma}^z$, 
the strength of the Rashba SO coupling through $\alpha^2$, and the momentum operator 
through the cross product $\hat{\bf p} \times {\bf z}$, allows one to differentiate 
symmetry properties of the two spin Hall accumulation components upon changing the
Rashba electric field (i.e., the sign of $\alpha$) or the
direction of the packet propagation: $S_z({\bf r})_{-\alpha} =
S_z({\bf r})_{\alpha}$ and $S_z({\bf r})_{- {\bf p}} = - S_z({\bf
r})_{\bf p}$ vs. $S_x({\bf r})_{-\alpha} = - S_x({\bf
r})_{\alpha}$ (due to opposite spin precession for $-\alpha$) and
$S_x({\bf r})_{-{\bf p}} = S_x({\bf r})_{\bf p}$. These features
are in full accord with  experimentally observed behavior of
the $S_z({\bf r})$ spin Hall accumulation under the inversion of the  
bias voltage,~\cite{wunderlich} as well as with the formal quantitative 
quantum transport analysis~\cite{accumulation} of the 
{\em nonequilibrium} spin accumulation induced by the flow of unpolarized 
charge  current through ballistic SO coupled two-probe nanostructures.

Finally, we note that $\alpha^2$ dependence of the transverse SO ``force''  
is incompatible with the $\alpha$-independent  (i.e., ``universal'') intrinsic spin 
Hall conductivity  $\sigma_{sH}=e/8\pi$ (describing the pure transverse spin Hall 
current $j_y^z = \sigma_{sH} E_x$ of the $z$-axis polarized spin in response to the 
longitudinally applied electric field $E_x$) of an infinite homogeneous Rashba spin-split 
2DEG in the clean limit, which has been obtained within various bulk transport approaches.~\cite{sinova,rashba_eq,zhang,wave_packet,niu}  On the  other hand, it supports the
picture of the SO coupling dependent spin Hall accumulations~\cite{accumulation} 
$S_z({\bf r})$,  $S_x({\bf r})$ and the corresponding spin Hall 
conductances~\cite{meso_hall} (describing the $z-$ and the $x$-component 
of the  nonequilibrium spin Hall current in the transverse leads attached 
at the lateral edges of the Rashba wire) of the {\em mesoscopic} spin Hall effect 
in confined structures.~\cite{ring_hall,meso_hall,meso_hall_1} By the
same token, the sign of the spin accumulation on the edges (i.e.,
whether the spin current flows to the right or to the  left in the
transverse direction~\cite{meso_hall}) cannot be determined from the 
properties~\cite{niu}  of $\sigma_{sH}$. Instead one has to take into account the 
strength of the SO coupling $\alpha$ and the size  of the  device in the units of the 
characteristic mesoscale $L_{\rm SO}$, as demonstrated by Figs.~\ref{fig:force}  
and ~\ref{fig:she_accumulation}. This requirement stems from the oscillatory  
character  of the  transverse SO ``force'' brought about by the  spin precession 
of the deflected spins in the effective magnetic field of the Rashba SO coupled wires of 
finite width.

We are grateful to S. Souma, S. Murakami, Q. Niu, and J. Sinova for insightful discussions and 
E. I. Rashba for enlightening criticism. Acknowledgment is made to the donors of the American 
Chemical Society Petroleum Research Fund for partial support of this research.

%********************references************************************************************************

%*****************************************************************

\end{document}